\documentclass{svjour2}                    

\smartqed  
\usepackage{graphicx}
\begin{document}

\title{Spherical surface models with directors
\thanks{This work was supported in part by a Grant-in-Aid for Scientific Research from Japan Society for the Promotion of Science.} 
}

\titlerunning{Spherical surface models with directors}        

\author{Hiroshi Koibuchi}


\institute{Hiroshi Koibuchi \at
Department of Mechanical and Systems Engineering, Ibaraki National College of Technology, Nakane 866, Hitachinaka, Ibaraki 312-8508, Japan
 \\
              \email{koibuchi@mech.ibaraki-ct.ac.jp}           
}

\date{Received: date / Accepted: date}

\maketitle

\begin{abstract}
A triangulated spherical surface model is numerically studied, and it is shown that the model undergoes phase transitions between the smooth phase and the collapsed phase. The model is defined by using a director field, which is assumed to have an interaction with a normal of the surface. The interaction between the directors and the surface maintains the surface shape. The director field is not defined within the two-dimensional differential geometry, and this is in sharp contrast to the conventional surface models, where the surface shape is maintained only by the curvature energies. We also show that the interaction makes the Nambu-Goto model well-defined, where the bond potential is given by the area of triangles; the Nambu-Goto model is well-known as an ill-defined one even when the conventional two-dimensional bending energy is included in the Hamiltonian.

\end{abstract}

\section{Introduction}\label{intro}
Shape of membranes sensitively changes depending on certain specific environmental conditions such as flow fields, gravity, and thermal fluctuations \cite{SEIFERT-LECTURE2004}. Transformations of the shape, as well as the surface fluctuations, are typical to soft materials such as biological membranes \cite{NELSON-SMMS2004}. Considerable efforts have been given so far to understand these phenomena through statistical mechanical treatments especially from the view point of phase transitions \cite{GK-SMMS2004,Bowick-PREP2001}.

Surface models for such phenomena are conventionally defined by using curvature Hamiltonians \cite{HELFRICH-1973,POLYAKOV-NPB1986,KLEINERT-PLB1986}. A well-known two-dimensional curvature energy is the so-called Helfrich and Polyakov Hamiltonian, which is rigorously defined by using the notions of two-dimensional differential geometry and plays a role for maintaining the surface shape \cite{Peliti-Leibler-PRL1985,DavidGuitter-EPL1988,PKN-PRL1988}. A linear bending energy in the compartmentalized surfaces can also maintain the surface shape \cite{KOIB-PRE-2007-2,KOIB-PRE-2007-3,KOIB-PLA-2007,KOIB-EPJB-2007-3,KOIB-JSTP-20067}. A unit tangential vector along the compartment defines the linear bending energy in those compartmentalized models  \cite{KOIB-PRE-2007-2,KOIB-PRE-2007-3,KOIB-PLA-2007,KOIB-EPJB-2007-3,KOIB-JSTP-20067}, while a unit normal vector of the triangles defines the two-dimensional bending energy of Helfrich and Polyakov in the conventional models \cite{KANTOR-NELSON-PRA1987,KOIB-PRE-2005,KOIB-NPB-2006}. Thus, we see that the mechanical strength of the surface is always provided by objects defined within the surface geometry such as the linear bending energy and the two-dimensional bending energy. 

On the other hand, director fields are crucial to understand phenomena such as the main transition in liquid crystals including Langmuir monolayer \cite{NELSON-SMMS2004}. The directors represent lipid molecules and can simply be described by the three-dimensional vectors on the surface. The chirality of membranes is also connected with the tilt of directors \cite{HelfrichProst-PRA1988,ZJX-PRL1990,SelMacSch-PRE1996,TuSeifert-PRE2007,NELSON-POWERS-PRL-1992,NELSON-POWERS-JPIIFR-1992}. The directors align to each others and become ordered at low temperature, while they become disordered at high temperature. It should also be noted that the directors, unlike the normal vectors of the surface, cannot be defined within the surface geometry.    

However, it is unclear at present whether the directors maintain the surface shape of membranes and what is the role of the director if it could maintain the surface shape. Therefore, interactions between the director fields and the surface are very interesting and still remain to be studied. We know that self-avoiding interactions can also maintain the surface shape against the collapse \cite{GREST-JPIF1991,BOWICK-TRAVESSET-EPJE2001,BCTT-PRL2001,Kroll-Gompper-JPF1993}, however, the collapsing transition of self-avoiding surfaces is very time consuming for numerical studies. 

In this paper, we investigate whether the surface shape can be maintained only by interactions between the directors and the surface. The problem we are interested in is whether or not the directors can provide mechanical strength to the surface for maintaining the shape. Two types of bond potentials are examined; the first is the conventional Gaussian bond potential and the second is the Nambu-Goto potential. It is well known that the surface model with the Nambu-Goto potential is ill-defined \cite{ADF-NPB-1985}, and the model is also ill-defined even when the curvature energy is included in the Hamiltonian. Therefore, we check whether the model with director is well-defined when the Nambu-Goto energy is assumed as the bond potential. Moreover, our interest focuses on the phase structure of the model if the surface shape is geometrically well-defined. In that case, it is also interesting to see whether or not the phase structure of the model is different from those of conventional surface models defined by the above mentioned curvature Hamiltonians. 

\section{Models}\label{model}
The triangulated sphere, where the models are defined, is constructed from the icosahedron and is identical with those in \cite{KOIB-PRE-2005}. By partitioning the icosahedron such that a bond of the icosahedron is split into $\ell$ pieces, we have a triangulated surface of size $N=10\ell^2\!+\!2$, in which $12$ vertices are of coordination number $q\!=\!5$, and the remaining $N\!-\!12$ vertices are of $q\!=\!6$. 

The models are defined statistical mechanically and hence by the following partition function:
\begin{equation} 
\label{Part-Func}
 Z = \sum_{\bf d} \int^\prime \prod _{i=1}^{N} d X_i \exp\left[-S(X,{\bf d})\right], \quad
 S(X,{\bf d})=S_1 + b S_2. 
\end{equation} 
The parameter $b$ is the microscopic bending rigidity and is of unit $kT$, where $k$ and $T$ are the Boltzmann constant and the temperature, respectively. $S(X,{\bf d})$ is the Hamiltonian, which is dependent on the variables $X$ and ${\bf d}$; $X$ represents the three dimensional position of vertices and ${\bf d}$ represents a three dimensional unit vector, which will be defined below. The symbol $\int^\prime \prod _{i=1}^{N} d X_i$ denotes that the center of mass of the surface is fixed in the integrations.  $S_1$ and $S_2$ are defined as follows:
\begin{equation}
\label{Disc-Eneg-S1S2-1} 
S_1=\sum_{(ij)} (X_i-X_j)^2, \quad
S_2=\sum_i \sum_{j(i)} [1-{\bf d}_i \cdot {\bf n}_{j(i)}],\qquad({\rm model}\; 1),
\end{equation} 
and
\begin{equation}
\label{Disc-Eneg-S1S2-2} 
S_1=\sum_{\it \Delta} A_{\it \Delta},  \quad 
S_2=\sum_i \sum_{j(i)} [1-{\bf d}_i \cdot {\bf n}_{j(i)}], \qquad\qquad ({\rm model}\; 2).
\end{equation} 
The bond potential $S_1$ of model 1 in Eq.(\ref{Disc-Eneg-S1S2-1}) is the Gaussian  bond potential, which is defined by the sum of bond length squares, while $S_1$ of model 2 in Eq.(\ref{Disc-Eneg-S1S2-2}) is called the Nambu-Goto energy, which is defined by the sum of the area of triangles ${\it \Delta}$. 

\begin{figure}[htb]
\centering
\includegraphics[width=4cm]{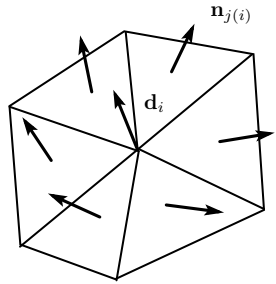}
\caption{A director field ${\bf d}_i$ at the vertex $i$, and the unit normal vectors ${\bf n}_{j(i)}$ which interact with ${\bf d}_i$.  } 
\label{fig-1}
\end{figure}
The symbol ${\bf d}_i $ in $S_2$ of Eqs.(\ref{Disc-Eneg-S1S2-1}) and (\ref{Disc-Eneg-S1S2-2}) is a three dimensional unit vector defined at the vertex $i$. We call ${\bf d}_i $ as the {\it director field} or simply as the {\it director}. The symbols  ${\bf n}_{j(i)} $ in $S_2$ is a unit normal vector of the triangle $j(i)$ surrounding the vertex $i$. The definition of $S_2$ in model 2 is identical to that in model 1. Since $S_2$ is very similar to the bending energy of the first model in \cite{KOIB-NPB-2006}, we call $S_2$ as the bending energy. The director field ${\bf d}_i $ and the unit normal vectors ${\bf n}_{j(i)} $ are shown in Fig.\ref{fig-1}.

The difference between model 1 and model 2 is seen only in the definitions of $S_1$. We should note that the Nambu-Goto energy $S_1$ of Eq.(\ref{Disc-Eneg-S1S2-2}) makes the model ill-defined if $S_2$ is given by the conventional curvature Hamiltonian such that $S_2=\sum_{(ij)} (1-{\bf n}_i \cdot {\bf n}_j)$. It should also be noted that the director energy $S_2$ in Eqs.(\ref{Disc-Eneg-S1S2-1}) and (\ref{Disc-Eneg-S1S2-2}) is not defined within the surface geometry because the director ${\bf d}_i $ is an external variable of the surface; it is not constructed without points outside the surface. For this reason $S_2$ is completely different from the bending energy of the first model in \cite{KOIB-NPB-2006}, although the expressions of $S_2$ in Eq.(\ref{Disc-Eneg-S1S2-1}) and the bending energy in \cite{KOIB-NPB-2006} are very similar to each other as mentioned above.  
 
We note that the models are symmetric under the transformations such that ${\bf n}\to -{\bf n}$ and ${\bf d}\to -{\bf d}$, where the normal vector ${\bf n}$ is chosen to have only one of the two-orientations of the surface. This symmetry implies the existence of the phase transition between two potential minima in the smooth phase. Although the orientation can not change from one to the other in the case of self-avoiding closed surfaces, it can change in our models, which are phantom. This symmetric property is identical to that of the conventional curvature surface model on the phantom spherical surface, where the model is symmetric under ${\bf n}\to -{\bf n}$. Therefore, the phase structure of the models in this paper is expected to have the same phase structure as the conventional curvature model.

It should also be noted that the bending energy in this paper is different from the so-called elastic energy $(1/2) k_t ({\bf n} \wedge {\bf d})^2$ of Helfrich in \cite{HELFRICH-1973}, where ${\bf n}$ is the normal of the surface and ${\bf d}$ is the average orientation of the molecules, and $k_t$ is an elastic modulus. Since $({\bf n} \wedge {\bf d})^2$ can also be written as $\sin^2\theta$ by using the angle $\theta$ between ${\bf n}$ and ${\bf d}$, then $({\bf n} \wedge {\bf d})^2$ appears to be equal to $1-{\bf d} \cdot{\bf n}$ in $S_2$ of Eqs.(\ref{Disc-Eneg-S1S2-1}) and (\ref{Disc-Eneg-S1S2-2}) at sufficiently small $\theta$. However, $\theta$ is not always constrained to be sufficiently small in this paper. In fact, the angle $\theta$ in this paper has values in $0\leq \theta \leq \pi$ while ${\bf d}$ in \cite{HELFRICH-1973} is allowed to have values only in $0\leq \theta \leq \pi/2$; this is because the variable ${\bf d}$ in \cite{HELFRICH-1973} is assumed to be on the unit half-sphere while ${\bf d}$ in this paper is assumed to be on the whole unit sphere.

\section{Monte Carlo technique}\label{MC-Techniques}
The dynamical variables $X$ and ${\bf d}$ of the models are integrated out and summed over in the partition function of Eq.(\ref{Part-Func}). The integrations and the summations can be performed by the canonical Metropolis Monte Carlo (MC) technique on the triangulated surfaces. The random three-dimensional shift of vertices $X\to X^\prime=X\!+\!\delta X$ is accepted with the probability ${\rm Min}[1,\exp(-\delta S)]$, where $\delta S=S({\rm new})\!-\!S({\rm old})$. The random vector $\delta X$ is chosen in a small sphere, whose radius is fixed at the beginning of the simulations so as to have $50\%$ acceptance rate. The variable ${\bf d}$ can also be updated in almost the same technique as that of $X$. The new position ${\bf d}^\prime$ is chosen on the unit sphere maintaining about $50\%$ acceptance rate. $N$ consecutive updates of $X$ and those of  ${\bf d}$ make one Monte Carlo sweep (MCS). 

The phase transitions are relatively strong rather than those in the conventional curvature models. For this reason, we concentrate on relatively small lattices of size up to $N\!=\!3612$ in model 1 and $N\!=\!5762$ in model 2. The transitions are hardly computed on large sized lattices, because the correlation time becomes longer and longer with increasing $N$.    

The total number of MCS after sufficiently large thermalization MCS is about $1\times10^8\sim 4\times10^8$ for the model 1 surfaces of size $N\!=\!812$ and $N\!=\!1212$,   $8\times10^8\sim 15\times10^8$  for those of $N\!=\!1962$, $N\!=\!2562$, and $N\!=\!3612$. About $5\times10^8$ MCS is performed for the model 2 surfaces of size $N\!=\!1692$, and $10\times10^8 \sim 16\times 10^8$ MCS for those of $N\!=\!2562$, $N\!=\!3612$, and $N\!=\!5762$.    

\section{Results}\label{Results}
\subsection{Snapshots, mean square size, and bond potentials}
\begin{figure}[htb]
\centering
\includegraphics[width=7.5cm]{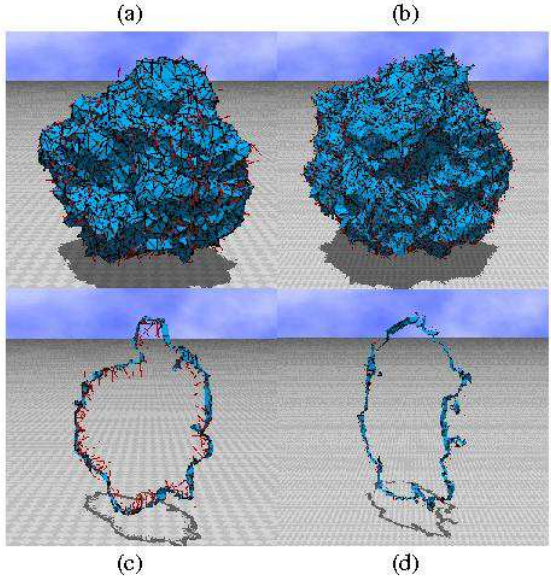}
\caption{(Color online) Snapshots of the surfaces of size $N\!=\!3612$ of (a) model 1 at $b\!=\!0.787$  and (b) model 2 at $b\!=\!0.8545$, (c) the surface section of (a), and (d) the surface section of (b). The mean square size is $X^2\!\simeq\!44$ in (a) and  $X^2\!\simeq\!331$ in (b). Pins or burs sticking out of the surfaces are the directors ${\bf d}_i$.  } 
\label{fig-2}
\end{figure}
Firstly, we show snapshots of surfaces and their surface sections of model 1 and model 2 in Figs. \ref{fig-2}(a) -- \ref{fig-2}(d). The surfaces are obtained in the smooth phase at the transition points $b\!=\!0.787$ and  $b\!=\!0.8545$ of model 1 and model 2, respectively. Small pins or burs seen on the surfaces represent the director fields ${\bf d}_i$, which are of unit length. The directors are hardly seen on the surface of model 2, because the surface size of model 2 in Fig. \ref{fig-2}(b) is quite larger than that of model 1 in Fig. \ref{fig-2}(a). From the surface section of Fig. \ref{fig-2}(c), we see that almost all directors turn inside the surface. This indicates that the orientation of the surface in Fig. \ref{fig-2}(a) is opposite to the initial one. The surface size can be characterized by the mean square size $X^2$ defined by
\begin{equation}
\label{X2}
X^2={1\over N} \sum_i \left(X_i-\bar X\right)^2, \quad \bar X={1\over N} \sum_i X_i,
\end{equation}
where the symbol $\bar X$ in Eq.(\ref{X2}) is the center of mass of the surface. In fact, we have $X^2\!\simeq\!44$ in Fig. \ref{fig-2}(a) and  $X^2\!\simeq\!331$ in Fig. \ref{fig-2}(b).
 
\begin{figure}[htb]
\centering
\includegraphics[width=10.5cm]{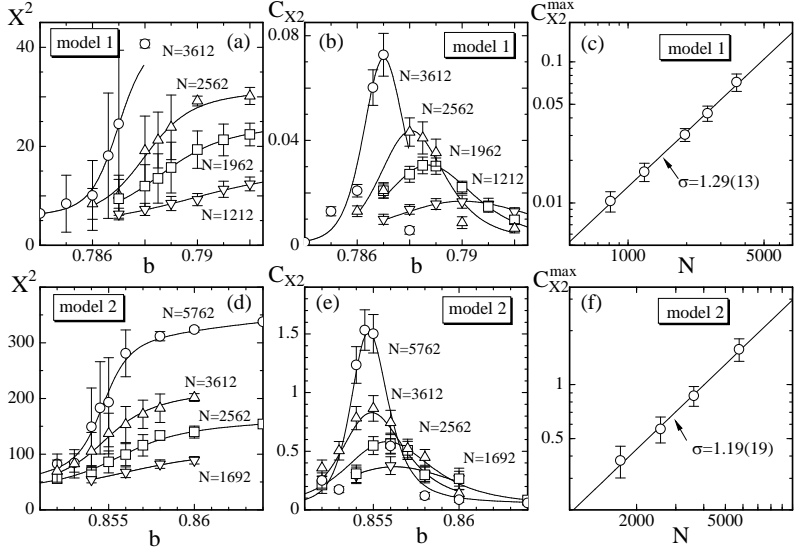}
\caption{(a) The mean square size $X^2$ vs. $b$ of model 1, (b) the variance $C_{X^2}$ vs. $b$  of model 1,  and  (c) a log-log plot of the peak values $C_{X^2}^{\rm max}$ vs. $N$ of model 1. (d), (e), (f) are those of model 2 corresponding to (a), (b), (c). The error bars in (a) and (d) are the standard errors, and those in (b), (c), (e), and (f) are the statistical errors. The solid curves are drawn by the multihistogram reweighting technique. } 
\label{fig-3}
\end{figure}
The surface shape can also be reflected in the mean square size $X^2$. Figures \ref{fig-3}(a) and \ref{fig-3}(d) show $X^2$ versus $b$ of model 1 and model 2, respectively. The solid curves in Figs. \ref{fig-3}(a), \ref{fig-3}(b), \ref{fig-3}(d), and \ref{fig-3}(e) are drawn by the multihistogram reweighting technique \cite{Janke-histogram-2002}. We see that $X^2$ grows larger and larger against $b$ with increasing $N$ in both models. This indicates a collapsing transition between the smooth phase and the collapsed phase, although no discontinuity is seen in $X^2$. To see the order of the transition, we plot the variance $C_{X^2}$ of $X^2$ in Figs. \ref{fig-3}(b) and  \ref{fig-3}(e), where  $C_{X^2}$ is defined by 
\begin{equation}
\label{CX2}
C_{X^2} \!=\! (1/ N) \langle \; \left( X^2 \!-\! \langle X^2 \rangle\right)^2\rangle.
\end{equation}
The peak values $C_{X^2}^{\rm max}$ are plotted against $N$ in Figs. \ref{fig-3}(c) and \ref{fig-3}(f) in a log-log scale. The straight lines in Figs. \ref{fig-3}(c) and \ref{fig-3}(f) are drawn by fitting the data to 
\begin{equation}
\label{scaling-CX2-N}
C_{X^2}^{\rm max}\sim N^\sigma,
\end{equation}
where $\sigma$ is a critical exponent of the transition. Thus, we have
\begin{eqnarray}
\label{exponent-CX2-N}
 &&\sigma=1.29\pm0.13,\quad ({\rm model \; 1}), \nonumber \\
 &&\sigma=1.19\pm0.19,\quad ({\rm model \; 2}).
\end{eqnarray}
The value of $\sigma$ of model 1 is slightly larger than $\sigma\!=\!1$ and the one of model 2 is almost identical to $\sigma\!=\!1$. Both results indicate that $C_{X^2} \to \infty$ in the limit of $N\to \infty$, and therefore the order of the collapsing transition is considered to be first order from the finite-size scaling theory \cite{PRIVMAN-WS-1989,BINDER-RPP-1997,BNB-NPB-1993}. 
 
\begin{figure}[htb]
\centering
\includegraphics[width=7.5cm]{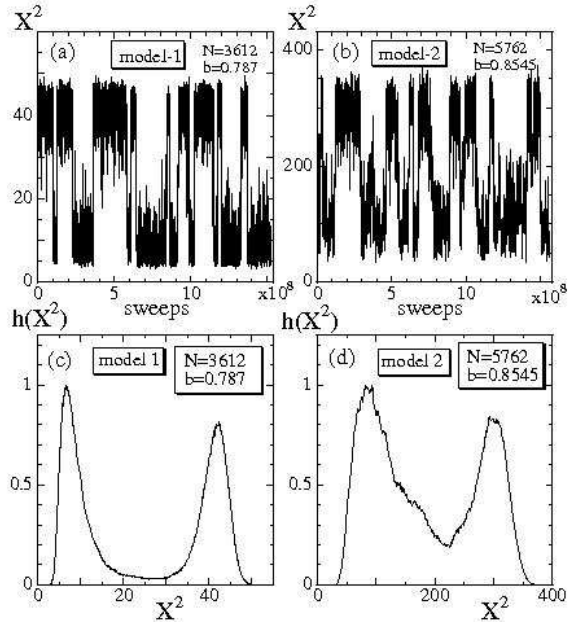}
\caption{ (a) The variation of $X^2$ against MCS obtained at the transition point $b\!=\!0.787$ on the surface of size $N\!=\!3612$ of model 1, (b) the variation at $b\!=\!0.8545$ on the surface of size $N\!=\!5762$ of model 2, and the normalized distribution (or histogram)  $h(X^2)$ of the variation $X^2$ of (c) model 1 and (d) model 2.  } 
\label{fig-4}
\end{figure}
The first-order nature of the transition can be seen more convincingly in the variation of $X^2$ against MCS at the transition point. Figures \ref{fig-4}(a) and \ref{fig-4}(b) show $X^2$ vs. MCS, which were respectively obtained at the transition point $b\!=\!0.787$ on the $N\!=\!3612$ surface of model 1 and at $b\!=\!0.8545$ on the $N\!=\!5762$ surface of model 2. Two distinct states are clearly seen in the variations of $X^2$; one is the smooth state and the other is the collapsed state. The first-order collapsing transition can also be confirmed from a double peak structure in the distribution (or histogram) $h(X^2)$ of the variation of $X^2$. Figures \ref{fig-4}(c) and \ref{fig-4}(d) shows $h(X^2)$ of model 1 and model 2, where the existence of the double peaks is clearly seen.  

\begin{figure}[htb]
\centering
\includegraphics[width=7.5cm]{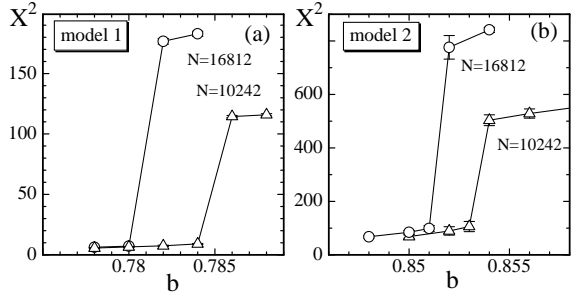}
\caption{$X^2$ vs. $b$ of (a) model 1 and  (b) model 2  on relatively large surfaces. The solid lines are drawn to guide the eyes. }
\label{fig-5}
\end{figure}

 We know that the curvature surface models undergo a first-order collapsing transition \cite{KOIB-PRE-2005,KOIB-NPB-2006}, therefore, the order of the transition remains unchanged if the curvature energy is replaced by the bending energy in this paper. However, the transitions in model 1 and model 2 seem rather strong than those of the curvature energy models. In fact, $X^2$ in the collapsed phase in Figs. \ref{fig-3}(a) and \ref{fig-3}(d) is almost independent of $N$, and therefore, the Hausdorff dimension of the surface in the collapsed phase is expected to be $H>3$, which is typical of strong transitions seen in phantom surface models such as the tensionless model \cite{KOIB-NPB-2006} and the intrinsic curvature models \cite{KOIB-PLA-2005,KOIB-EPJB-2004}. Figures \ref{fig-5}(a) and \ref{fig-5}(b) show  $X^2$ vs. $b$ of model 1 and  model 2 on the $N\!=\!10242$ and $N\!=\!16812$ surfaces, which are relatively larger than those in Figs. \ref{fig-3}(a) and \ref{fig-3}(d). The total number of MCS is about $2\times10^8\sim 3\times10^8$ in the smooth phase and $1\times10^8$ in the collapsed phase in both models, and these numbers seems insufficient for such large sized surfaces. However, we understand from Figs. \ref{fig-5}(a) and \ref{fig-5}(b) that the transition is very strong.  

\begin{figure}[htb]
\centering
\includegraphics[width=7.5cm]{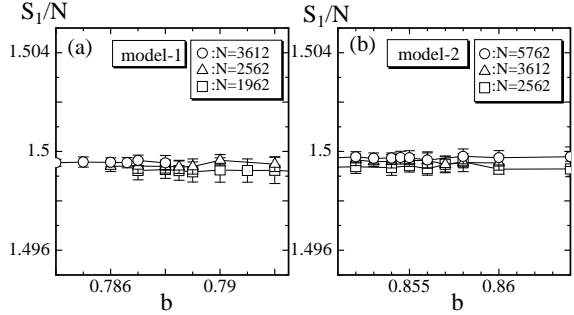}
\caption{(a) The Gaussian bond potential $S_1/N$ vs. $b$ of model 1 and  (b) the Nambu-Goto bond potential $S_1/N$ vs. $b$ of model 2. The expected relation $S_1/N\!\simeq\! 3/2$ is satisfied in both models. }
\label{fig-6}
\end{figure}
Finally in this subsection, the Gaussian bond potential and the Nambu-Goto bond potential $S_1/N$ are shown in Figs. \ref{fig-6}(a) and \ref{fig-6}(b), respectively, and we find that the simulations are successfully performed. From the scale invariance of the partition function, we have $S_1/N\!=\!3(N-1)/(2N)\!\simeq\!3/2$ if the center of mass of the surface is fixed. We find from Figs. \ref{fig-6}(a) and \ref{fig-6}(b) that the expected relation $S_1/N\!\simeq\! 3/2$ is satisfied.

\subsection{Bending energy and surface fluctuations}
\begin{figure}[hbt]
\centering
\includegraphics[width=10.5cm]{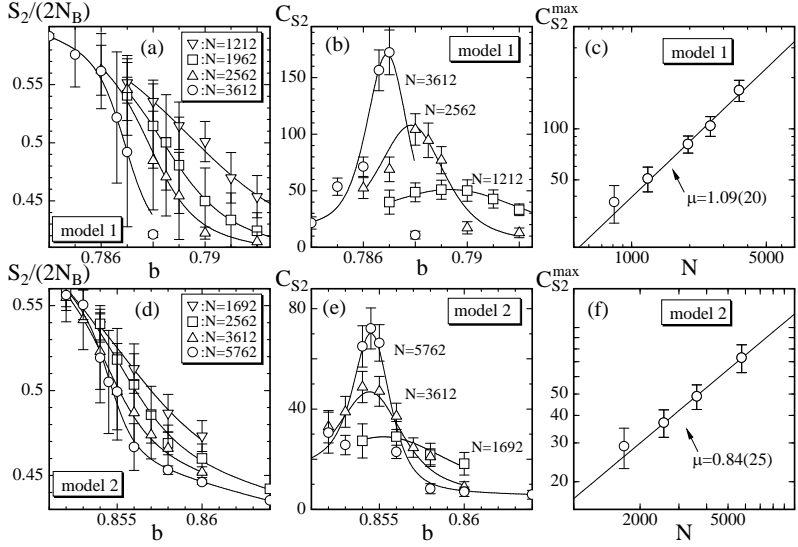}
\caption{(a) The bending energy $S_2/(2N_B)$ vs. $b$ of model 1, (b) the variance $C_{S_2}$ vs. $b$ of model 1, and (c) a log-log plot of the peak values $C_{S_2}^{\rm max}$ vs. $N$ of model 1. (d), (e), (f) are those of model 2 corresponding to (a), (b), (c). The error bars in (a) and (d) are the standard errors, and those in (b), (c), (e), and (f) are the statistical errors.  The solid curves are drawn by the multihistogram reweighting technique.} 
\label{fig-7}
\end{figure}
Figures \ref{fig-7}(a) and \ref{fig-7}(d) show the bending energy $S_2/(2N_B)$ versus $b$ obtained in model 1 and model 2. The summation $\sum_i\sum_{j(i)}$ in the definition of $S_2$ in Eq. (\ref{Disc-Eneg-S1S2-1}) or in Eq. (\ref{Disc-Eneg-S1S2-2}) gives $\sum_i\sum_{j(i)}1\!=\!2N_B$, where $N_B$ is the total number of bonds. This is the reason why $S_2$ is divided by $2N_B$ in Figs. \ref{fig-7}(a) and \ref{fig-7}(d). Discontinuous changes of $S_2/(2N_B)$ are not so apparent in the figures just like $X^2$ in Figs. \ref{fig-3}(a) and \ref{fig-3}(d). The variance of $S_2$ is given by $C_{S_2} \!=\! (1/ N) \langle \; \left( S_2 \!-\! \langle S_2 \rangle\right)^2\rangle$ and is shown in Figs. \ref{fig-7}(b) and \ref{fig-7}(e). Peaks are seen in $C_{S_2}$ and indicate that the models undergo the transition of surface fluctuations. We show the peak values $C_{S_2}^{\rm max}$ vs. $N$ in Figs. \ref{fig-7}(c) and \ref{fig-7}(f) in a log-log scale. The scaling behavior
is observed such that $C_{S_2}^{\rm max}\sim N^\mu$ in both models, and we have
\begin{eqnarray}
\label{exponent-CS2-N}
&&\mu=1.09\pm0.20,\quad ({\rm model \; 1}), \nonumber \\
&&\mu=0.84\pm0.25,\quad ({\rm model \; 2}),
\end{eqnarray}
where the fitting was done by using the largest four data in Fig. \ref{fig-7}(c) and  the largest three data in Fig. \ref{fig-7}(f). Thus, we see that $\mu\simeq 1$ in model 1, and therefore the order of the transition of surface fluctuation is considered to be of first order. In the case of model 2, $\mu$ is considered to be $\mu\simeq 1$ within the error, and therefore the result is consistent with the first-order transition. We should note that the transition point $b_c(N)$, where $C_{S_2}$ has the peak, is almost identical to that for the collapsing transition, which is seen in Fig. \ref{fig-3}. Thus, the transition of surface fluctuations is considered to occur at the same transition point of the collapsing transition in each model.

\begin{figure}[htb]
\centering
\includegraphics[width=7.5cm]{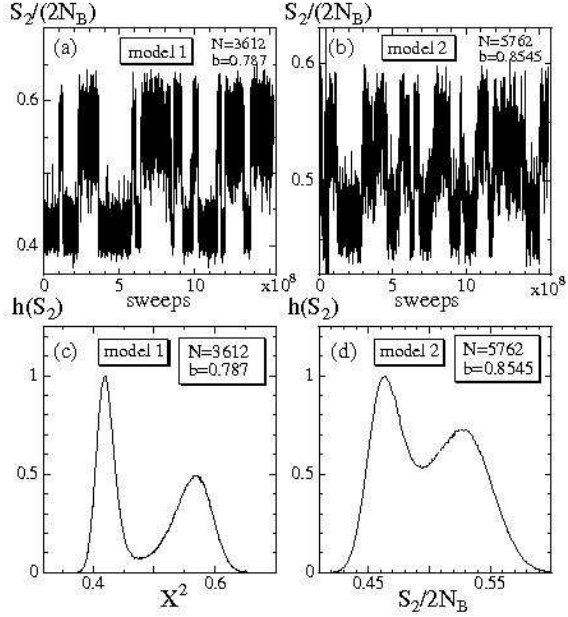}
\caption{(a) The variation of $S_2/(2N_B)$ against MCS obtained at the transition point $b\!=\!0.787$ on the surface of size $N\!=\!3612$ of model 1, (b) the variation at $b\!=\!0.8545$ on the surface of size $N\!=\!5762$ of model 2, and the normalized histogram $h(S_2)$ of the variation $S_2/(2N_B)$ of (c) model 1 and (d) model 2.  } 
\label{fig-8}
\end{figure}
Figures \ref{fig-8}(a) and \ref{fig-8}(b) show the variations of $S_2/(2N_B)$ against MCS at the transition points  of model 1 and model 2. $S_2/(2N_B)$ in Fig. \ref{fig-8}(a) is obtained at $b\!=\!0.787$ on the $N\!=\!3612$ surface, while $S_2/(2N_B)$ in Fig. \ref{fig-8}(b) is at $b\!=\!0.8545$ on the $N\!=\!5762$ surface. The histograms $h(S_2)$ of model 1 and mode 2 are shown in Figs. \ref{fig-8}(c) and \ref{fig-8}(d), respectively. We clearly see a double peak structure in both $h(S_2)$. Thus, we have confirmed more clearly from the variations of $S_2/(2N_B)$ and the histogram $h(S_2)$ that the order of the transition is of first order. 

\begin{figure}[htb]
\centering
\includegraphics[width=10.5cm]{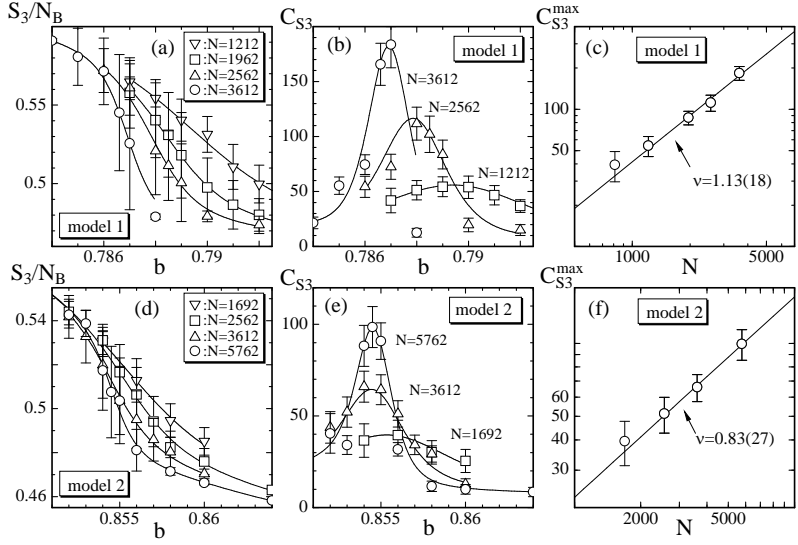}
\caption{(a) The bending energy $S_3/N_B$ vs. $b$ of model 1, (b) the variance $C_{S_3}$ vs. $b$  of model 1,  and  (c) a log-log plot of the peak values $C_{S_3}^{\rm max}$ vs. $N$ of model 1. (d), (e), (f) are those of model 2 corresponding to (a), (b), (c). The error bars in (a) and (d) are the standard errors, and those in (b), (c), (e), and (f) are the statistical errors. The solid curves are drawn by the multihistogram reweighting technique.} 
\label{fig-9}
\end{figure}
The surface fluctuations are assumed to be reflected in the bending energy $S_2/(2N_B)$. However, $S_2$ includes the director field ${\bf d}_i$, which is not directly connected to the surface fluctuations. Therefore, it is interesting to see the bending energy $S_3$ defined by $S_3=\sum_{(ij)} (1\!-\!{\bf n}_i\cdot {\bf n}_j)$, where ${\bf n}_i$ and ${\bf n}_j$ are unit normal vectors of the triangles $i$ and $j$, which have a common bond. Although $S_3$ is not included in the Hamiltonian, $S_3$ is considered to reflect the surface fluctuations.

In order to see the transition of surface fluctuations, we plot $S_3/N_B$ in Figs. \ref{fig-9}(a) and \ref{fig-9}(d). The corresponding variance $C_{S_3}$ defined by $C_{S_3} \!=\! (1/ N) \langle \; \left( S_3 \!-\! \langle S_3 \rangle\right)^2\rangle$ is also plotted in Figs. \ref{fig-9}(b) and \ref{fig-9}(e) against $b$. The peak values $C_{S_3}^{\rm max}$ against $N$ are shown in  Figs. \ref{fig-9}(c) and \ref{fig-9}(f) in the log-log scale. We find that the behavior of $S_3/N_B$ is almost identical to that of the bending energy $S_2/(2N_B)$ in Figs. \ref{fig-7}(a) and \ref{fig-7}(d), and that the shape of the curves of $C_{S_3}$ is almost identical to that of $C_{S_2}$ in Figs. \ref{fig-7}(b) and \ref{fig-7}(e). The exponent $\nu$ defined by $C_{S_3}^{\rm max}\sim N^\nu$ is obtained by fitting the data in Figs. \ref{fig-9}(c) and \ref{fig-9}(f), and we have $\nu\!=\!1.13(18)$ and $\nu\!=\!0.83(27)$ for model 1 and model 2, respectively. Both of $\nu$ are consistent with the values of $\mu$ in Eq. (\ref{exponent-CS2-N}) as expected. 

\begin{figure}[htb]
\centering
\includegraphics[width=7.5cm]{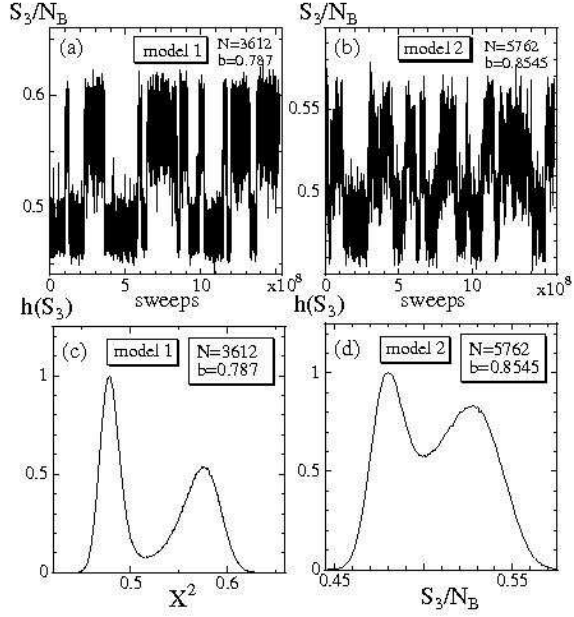}
\caption{(a) The variation of $S_3/N_B$ against MCS obtained at the transition point $b\!=\!0.787$ on the surface of size $N\!=\!3612$ of model 1, (b) the variation at $b\!=\!0.8545$ on the surface of size $N\!=\!5762$ of model 2, and the normalized histogram $h(S_3)$ of the variation $S_3/N_B$ of (c) model 1 and (d) model 2. } 
\label{fig-10}
\end{figure}
The variation of $S_3/N_B$ vs. MCS is shown in Figs. \ref{fig-10}(a) and \ref{fig-10}(b). Both of which are obtained at the same transition points where the variations of $S_2/(2N_B)$ in Figs. \ref{fig-8}(a) and \ref{fig-8}(b) are obtained. We see that $S_3/N_B$ changes between two different states corresponding to the smooth phase and the collapsed phase in both models, and the behaviors of the variation of  $S_3/N_B$ are almost identical with those of $S_2/(2N_B)$ shown in Figs. \ref{fig-8}(a) and \ref{fig-8}(b). This confirms that the transition of surface fluctuations is of first-order. Figures  \ref{fig-10}(c) and \ref{fig-10}(d) show the normalized histogram $h(S_3)$ of the variation $S_3/N_B$ of model 1 and model 2, and we see in $h(S_3)$ a double peak structure, which corresponds to that in $h(S_2)$ in Figs. \ref{fig-8}(c) and \ref{fig-8}(d) and is consistent with the first-order transition of surface fluctuations. 

The correlation energy of the directors is defined by $S_4=\sum_{(ij)} (1\!-\!{\bf d}_i\cdot {\bf d}_j)$, where $\sum_{(ij)}$ denotes the sum over  bond $(ij)$ connecting the vertices $i$ and $j$, and thus $\sum_{(ij)}1\!=\!N_B$. We checked whether or not the transition of surface fluctuations is reflected in $S_4$, and  we reconfirmed from $S_4/N_B$ and the variance $C_{S_4}$ that the transition of surface fluctuations is of first-order. Although no figure of these quantities is shown, the behaviors of these quantities are almost identical with those of $S_2$ in Figs. \ref{fig-7} and \ref{fig-8} and those of $S_3$ in  Figs. \ref{fig-9} and \ref{fig-10}. The directors ${\bf d}_i$ become parallel to each other and normal to the surface if the surface is sufficiently smooth, and therefore ${\bf d}_i$ are expected to be strongly correlated to each other in the smooth phase. On the contrary, ${\bf d}_i$ are naturally expected to be uncorrelated on the fluctuated surfaces. Therefore, the transition of surface fluctuations is also reflected in $S_4/N_B$. 

\subsection{Scaling at $b>b_c$}
\begin{figure}[htb]
\centering
\includegraphics[width=7.5cm]{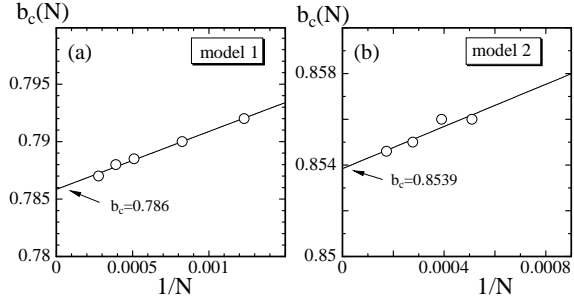}
\caption{The transition point $b_c(N)$ vs. $1/N$ of (a) model 1 and (b) model 2.  $b_c(N)$ are obtained from the peaks of $C_{X^2}$ in both models. The straight lines are drawn by fitting the data linearly with respect to $1/N$. } 
\label{fig-11}
\end{figure}
The transition point $b_c$ is obtained from $b_c(N)$ in the limit of $N\!\to\! \infty$, where  $b_c(N)$ is evaluated from the peaks of $C_{X^2}$ in Figs. \ref{fig-3}(b) and \ref{fig-3}(e) or of 
$C_{S_2}$ in Figs. \ref{fig-7}(b) and \ref{fig-7}(e). Figures \ref{fig-11}(a) and \ref{fig-11}(b) show $b_c(N)$ vs. $1/N$ in the linear scale, where $b_c(N)$ is obtained from the peaks of $C_{X^2}$ in Figs. \ref{fig-3}(b) and \ref{fig-3}(e). We have seen no clear difference between the transition point of the collapsing transition and that of the transition of surface fluctuations. For this reason, we show only $b_c(N)$ obtained from the peaks of $C_{X^2}$. The straight lines are obtained by assuming that $b_c(N)$ is proportional to $1/N$; $b_c(N)\!=\!b_c\!+\!a (1/N)$, where the parameter $b_c$ is the transition point in the thermodynamic limit of the models. Thus, we have $b_c\!=\!0.786$ for model 1 and $b_c\!=\!0.8539$ for model 2. We should note that the values of $b_c$ are inconsistent with the transition points expected from $X^2$ in Figs. \ref{fig-5}(a) and \ref{fig-5}(b), and the reason of this seems that the total number of MCS in the smooth phase was insufficient for the simulations of those large sized surfaces.

\begin{figure}[htb]
\centering
\includegraphics[width=7.5cm]{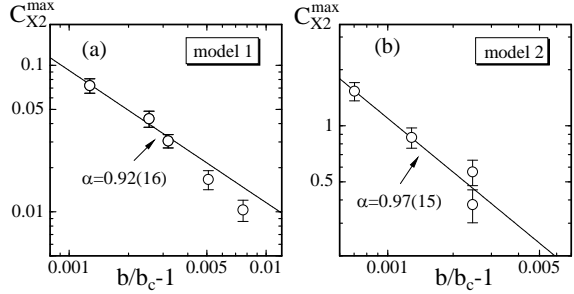}
\caption{The peak values $C_{X^2}^{\rm max}$ vs. $b/b_c\!-\!1$ of (a) model 1 and  (b) model 2 in a log-log scale, where $b_c\!=\!0.786$ in (a) and  $b_c\!=\!0.8539$ in (b). } 
\label{fig-12}
\end{figure}
The peak values $C_{X^2}^{\rm max}$ shown in Figs. \ref{fig-3}(b) and \ref{fig-3}(e) are plotted against $b/b_c\!-\!1$ in  Figs. \ref{fig-12}(a) and \ref{fig-12}(b) in a log-log scale, where $b_c\!=\!0.786$ and $b_c\!=\!0.8539$ are assumed. The straight line is the fitted one of the data such that $C_{X_2}^{\rm max}\!\sim\!(b/b_c\!-\!1)^{-\alpha}$ with exponents $\alpha\!=\!0.92\pm0.16$ and  $\alpha\!=\!0.97\pm0.15$. These represent a conventional scaling property of $C_{X_2}^{\rm max}$ at $b>b_c$ and indicate that both models undergo the first-order collapsing transition. The results are also consistent with the values in Eq. (\ref{exponent-CX2-N}). 

\begin{figure}[htb]
\centering
\includegraphics[width=7.5cm]{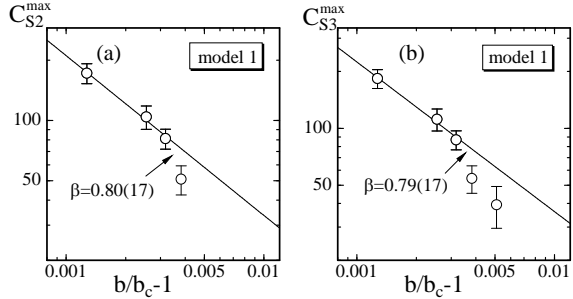}
\caption{(a) The peak values $C_{S_2}^{\rm max}$  vs. $b/b_c\!-\!1$ of model 1 and  (b) the peak values $C_{S_3}^{\rm max}$  vs. $b/b_c\!-\!1$ of model 1 in a log-log scale, where $b_c\!=\!0.786$ in (a) and  (b).} 
\label{fig-13}
\end{figure}
$C_{S_2}^{\rm max}$ and $C_{S_3}^{\rm max}$ of model 1 are also shown in Figs. \ref{fig-13}(a) and \ref{fig-13}(b) against $b/b_c\!-\!1$. The straight lines are drawn by fitting the largest three data in both of the figures.  Thus, we confirm that the conventional scaling properties on the variances of $S_2$ and $S_3$ are almost consistent with the fact that model 1 undergo a first-order transition of surface fluctuations. However, in the case of model 2 we can not always see the scaling property of $C_{S_2}^{\rm max}$ and $C_{S_3}^{\rm max}$ in contrast to the case of model 1 in Figs. \ref{fig-13}(a) and \ref{fig-13}(b). The reason of this seems only due to the low statistics of the simulations for model 2; the peak positions $b_c(N)$ corresponding to $C_{S_2}^{\rm max}$ and $C_{S_3}^{\rm max}$ in  Figs. \ref{fig-7}(c) and \ref{fig-9}(c) are not always consistent with the plots for this scaling. 
  
\section{Summary and Conclusion}\label{Conclusion}
To summarize, we have investigated a possible mechanism that the directors maintain the surface shape of membranes by using a spherical phantom surface model, which has no curvature Hamiltonian. The directors in the model are analogues of lipid molecules or some external molecules in membranes, whose shape is considered to be crucially dependent on the three-dimensional structure of the molecules. We focused our attentions on an interaction between the directors and the surface, and the phase structure of the model was studied by using the canonical MC simulation technique. We should note that the Helfrich Hamiltonian includes the mean curvature squared term and the Gaussian curvature term. The bending energy in this paper, as well as the conventional bending energy of the type $1\!-\!{\bf n}\cdot {\bf n}$, corresponds to the mean curvature squared term in the Helfrich Hamiltonian. The Gaussian curvature term is eliminated from the Hamiltonian of the model in this paper. This term seems relevant to the shape of such surfaces with holes, however, it can be neglected in the case of closed surfaces because of the Gauss-Bonnet theorem \cite{FD-SMMS2004}.

Two-types of bond potentials are assumed in the model of this paper: The first is the standard Gaussian bond potential, which is included in the Hamiltonian of model 1, and the second is the so-called Nambu-Goto energy, which is defined by the area of the triangles and included in the Hamiltonian of model 2. Both model 1 and model 2 have the bending energy, which describes the interaction between the directors and the surface. The shape of surface is maintained by this interaction.

We found that both models undergo a first-order collapsing transition between the smooth phase and the collapsed phase. Moreover, a first-order transition of surface fluctuations occurs in the models at the same transition point of the collapsing transition.  

One remarkable result is that the model is well-defined even when the Hamiltonian includes the Nambu-Goto energy as the bond potential. This is confirmed from the numerical results of model 2. In the case when the Nambu-Goto energy is included in the Hamiltonian as the bond potential, the conventional curvature surface model becomes ill-defined \cite{ADF-NPB-1985}.  
   
The phase transitions seen in the models in this paper are relatively strong compared to those of the conventional models, although the order of the transitions is of first-order and hence identical to those of the conventional models. In fact, the collapsed surface at the transition point is completely collapsed in the models, while the collapsed surface of the conventional model is relatively swollen at the transition point. Consequently, the Hausdorff dimension $H$ at the collapsed surface is greater than the physical bound, i.e. $H>3$, in the models of this paper. This is in sharp contrast with the fact $H<3$ in the collapsed phase at the transition point of the conventional curvature surface model \cite{KOIB-PRE-2005}.  

We comment on the difference between the bending energy $1\!-\!{\bf d} \cdot{\bf n}$ in this paper and the elastic energy $({\bf n} \wedge {\bf d})^2$ in \cite{HELFRICH-1973}. The variable ${\bf d}$ in \cite{HELFRICH-1973} has values on the unit half-sphere while ${\bf d}$ in this paper has values on the whole unit sphere. Therefore, the energy $1\!-\!{\bf d} \cdot{\bf n}$ in this paper is different from  $({\bf n} \wedge {\bf d})^2$ in \cite{HELFRICH-1973}, although both energies are almost equal to each other on sufficiently smooth surfaces. 

From the numerical results obtained in this paper, we conclude that the surface shape of membranes can be maintained by non-surface geometric object such as the directors, which interact with the surface. Important point to note is that the directors are the external variables of the surface. This implies that the directors are not always identified with the lipid molecules and allows us to speculate as follows: If some external objects could be embedded in the membrane so as to have the assumed interaction with the membrane constituents, the surface shape can be controlled.

 It is interesting to study the phase structure of the fluid surface model with the directors. Correlations between the directors can be assumed as an energy term for maintaining the surface shape; we expect that the surface shape is maintained by the correlation energy.

This work is supported in part by a Grant-in-Aid for Scientific Research from Japan Society for the Promotion of Science.  




\begin{thebibliography}{00}

\bibitem{SEIFERT-LECTURE2004}
U. Seifert, Fluid Vesicles, in {Lecture Notes: Physics Meets Biology. From Soft Matter to Cell Biology.}, 35th Spring Scool, Institute of Solid State Research, Forschungszentrum J${\ddot {\rm u}}$lich (2004).


\bibitem{NELSON-SMMS2004}
D. Nelson, in {Statistical Mechanics of Membranes and Surfaces, Second Edition}, edited by  D. Nelson, T.Piran, and S.Weinberg, (World Scientific, 2004), p.1. 

\bibitem{GK-SMMS2004}
G. Gompper, and D.M. Kroll, in {Statistical Mechanics of Membranes and Surfaces, Second Edition}, edited by  D. Nelson, T.Piran, and S.Weinberg, (World Scientific, 2004), p.359. 


\bibitem{Bowick-PREP2001}
 M. Bowick and A. Travesset, Phys. Rep. {\bf 344} (2001) 255.


\bibitem{HELFRICH-1973}
 W. Helfrich, Z. Naturforsch, {\bf 28}c (1973) 693.

\bibitem{POLYAKOV-NPB1986}
 A.M. Polyakov, Nucl. Phys. B {\bf 268} (1986) 406.

\bibitem{KLEINERT-PLB1986}
 H. Kleinert, Phys. Lett. B {\bf 174} (1986) 335.

\bibitem{Peliti-Leibler-PRL1985}
 L. Peliti and S. Leibler, Phys. Rev. Lett. {\bf 54} (15)  (1985) 1690.

\bibitem{DavidGuitter-EPL1988}
 F. David and E. Guitter, Europhys. Lett,  {\bf 5} (8)  (1988) 709.

\bibitem{PKN-PRL1988}
M. Paczuski, M. Kardar, and D. R. Nelson, Phys. Rev. Lett. {\bf 60}, (1988)  2638.

\bibitem{KOIB-PRE-2007-2}
 H. Koibuchi, Phys. Rev. E, {\bf 75}, (2007) 051115. 

\bibitem{KOIB-PRE-2007-3}
 H. Koibuchi, Phys. Rev. E, {\bf 76}, (2007) 061105. 

\bibitem{KOIB-PLA-2007}
H. Koibuchi, Phys. Lett. A {\bf 371}, (2007) 278.

\bibitem{KOIB-EPJB-2007-3}
H. Koibuchi, Euro. Phys. J. B {\bf 59}, (2007) 405.

\bibitem{KOIB-JSTP-20067}
 H. Koibuchi, J. Stat. Phys. {\bf 127}, (2006) 457; {\bf 129}, (2007) 605. 


\bibitem{KANTOR-NELSON-PRA1987}
 Y. Kantor and  D.R. Nelson, Phys. Rev. A {\bf 36}  (1987) 4020.

\bibitem{KOIB-PRE-2005}
 H. Koibuchi and T. Kuwahata, Phys. Rev. E, {\bf 72}, (2005) 026124. 

\bibitem{KOIB-NPB-2006}
 I. Endo and H. Koibuchi, Nucl. Phys. B {\bf 732} [FS], (2006) 426. 

\bibitem{HelfrichProst-PRA1988} W. Helfrich and J. Prost, Phys. Rev. A  \textbf{38}, 3065 (1988).

\bibitem{ZJX-PRL1990} Ou-Yang Zhong-can and Liu Ji-xing, Phys. Rev. Lett. \textbf{65}, 1679 (1990).

\bibitem{SelMacSch-PRE1996} J. V. Selinger, F. C. MacKintosh and J. M. Schnur, Phys. Rev. E \textbf{53}, 3804 (1996).

\bibitem{TuSeifert-PRE2007} Z. C. Tu and U. Seifert, Phys. Rev. E \textbf{76}, 031603 (2007).

\bibitem{NELSON-POWERS-PRL-1992}
P. Nelson and T. Powers, Phys. Rev. Lett. {\bf 69} (1992) 3409.

\bibitem{NELSON-POWERS-JPIIFR-1992}
P. Nelson and T. Powers, J. Phys. II France {\bf 3} (1993) 1535. 


\bibitem{GREST-JPIF1991}
G. Grest, J. Phys. I (France) {\bf 1}, 1695 (1991).

\bibitem{BOWICK-TRAVESSET-EPJE2001}
M. Bowick and A. Travesset,  Eur. Phys. J. E {\bf 5}, 149 (2001).

\bibitem{BCTT-PRL2001}
M. Bowick, A. Cacciuto, G. Thorleifsson, and  A. Travesset, Phys. Rev. Lett. {\bf 87}, 148103 (2001).

\bibitem{Kroll-Gompper-JPF1993}
D.M. Kroll and G. Gompper, J. Phys. I France {\bf 3}, 1131 (1993).

\bibitem{ADF-NPB-1985}
J. Ambjorn, B. Durhuus and J. Frohlich, Nucl. Phys. B {\bf 257}, 433 (1985).


\bibitem{Janke-histogram-2002}
Wolfhard Janke, {Histograms and All That},
In: Computer Simulations of Surfaces and Interfaces, NATO Science Series, II. Mathematics, Physics and Chemistry - Vol. 114, Proceedings of the NATO Advanced Study Institute, Albena, Bulgaria, 9 - 20 September 2002, edited by B. Dunweg, D.P. Landau, and A.I. Milchev (Kluwer, Dordrecht, 2003), pp. 137 - 157. 

\bibitem{PRIVMAN-WS-1989}
V. Privman, {Finite-Size Scaling Theory}, In: {Finite Size Scaling and Numerical Simulation of Statistical Systems}, V. Privman, Eds. (World Scientific, 1989) p.1.

\bibitem{BINDER-RPP-1997}
K. Binder, Applications of Monte Carlo methods to statistical physics,  Reports on Progress in Physics \textbf{60}, 487 - 559 (1997).

\bibitem{BNB-NPB-1993}
A. Billoire, T. Neuhaus, B. Berg, Nucl.Phys. B \textbf{396}, 779 (1993).

\bibitem{BINDER-ZFPB-1981}
K. Binder, Z. Phys. B \textbf{43}, 119 (1981).

\bibitem{KOIB-PLA-2005}
M.Igawa, H.Koibuchi, and M.Yamada, Phys. Lett. A {\bf 338}, 433 (2005).

\bibitem{KOIB-EPJB-2004}
H.Koibuchi, N.Kusano, A.Nidaira, Z.Sasaki, and K.Suzuki, Euro. Phys. J. B {\bf 42}, 561 (2004).

\bibitem{FD-SMMS2004}
F. David, in {Statistical Mechanics of Membranes and Surfaces, Second Edition}, edited by  D. Nelson, T.Piran, and S.Weinberg, (World Scientific, 2004), p.149. 



\end{thebibliography}
\end{document}